\documentclass{elsart}
\usepackage{epsbox}
\begin{document}
\unitlength 1mm
\begin{frontmatter}
\title{Explicit conversion from the Casimir force\\
       to Planck's law of radiation}
\author{Kenji Fukushima\thanksref{fuku}} and
\author{Koichi Ohta}
\thanks[fuku]{Corresponding author.\\
    {\it E-mail address:} fuku@nt1.c.u-tokyo.ac.jp}
\address{Institute of Physics, University of Tokyo, 3-8-1 Komaba,
         Meguro-ku, Tokyo 153-8902, Japan}

\begin{abstract}
The Casimir force has its origin in finite modification of the
infinite zero-point energy induced by a specific boundary condition
for the spatial configuration. In terms of the imaginary-time
formalism at finite temperature, the root of Planck's law of radiation
can be traced back to finite modification of the infinite vacuum
energy induced by the periodic boundary condition in the temporal
direction. We give the explicit conversion from the Casimir force to
Planck's law of radiation, which shows the apparent correspondence
between the system bounded by parallel conducting plates and the
thermodynamic system. The temperature inversion symmetry and the
duality relation in the thermodynamics are also discussed. We conclude
that the {\it effective temperature} characterized by the spatial
extension should no longer be regarded as genuine temperature.
\end{abstract}
\begin{keyword}
Casimir force;
Planck's law of radiation;
temperature inversion symmetry;
duality;
thermodynamic functions
\end{keyword}
\end{frontmatter}

The history of physics has passed through a number of memorable
break-through points. Unacceptable as they might have looked at a
first glance, many apparently unusual concepts permeate among us today
after tremendous efforts. The discovery of celebrated {\em Planck's
law of radiation} by Planck \cite{pla00} in 1900 and the recognition
of the {\em Casimir force} by Casimir \cite{cas48} in 1948 should, of
course, be enumerated as monumental achievements. In the language of
the imaginary-time formalism of the finite-temperature field theory
\cite{kap89}, Planck's formula can be perceived as the energy
associated with the assembly of an infinite number of free oscillators
with the periodic boundary condition in the temporal (thermal)
direction. The phrase ``an infinite number of oscillators'' would
remind us of the Casimir force that arises from finite discrepancy of
the zero-point energy in the presence of a constraining boundary
condition. The precise measurement of the Casimir force demonstrated
in 1997 \cite{lam97} confirmed the Casimir effect experimentally, even
though no attempt using parallel plates results in successful
measurement so far yet.

These two phenomena are closely related to each other by the
$\mathrm{O}(4)$ symmetry of the Euclidean space-time. Thus writing the
thermodynamic functions at finite temperature $T$ and extension $l$ in
terms of a dimensionless parameter $\xi=Tl$, we can readily observe
the symmetry under the exchange $\xi\leftrightarrow1/4\xi$, which was
first noted by Brown and Maclay for the scaled free energy
\cite{bro69}. Santos and Tort have recently shown the extended
symmetry in the system confined in a conducting rectangular cavity
\cite{san00}. A concise overview on this topic is available from
Ref.\ \cite{rav89}. Many authors are still paying attention to this
symmetric property, that is named {\em temperature inversion symmetry}.
This property is a typical manifestation of the equivalence under the
exchange between the role of $\beta=\hbar c/k_{\mathrm{B}}T$ and that
of $l$. The numerical coefficient is required to compensate for the
difference of the boundary conditions: the temporal direction is
periodic while the spatial one is fixed. This means that the spatial
counterpart of the Matsubara frequency is half smaller. It is most
often the case that authors confirmed the realization of the
temperature inversion symmetry simply by looking over the resultant
expressions with some cutoff scheme at finite temperature and
extension. In this paper we will demonstrate the explicit conversion
to the same form as Planck's law of radiation beginning with the
definition of the Casimir force, which is the simplest example of the
temperature inversion symmetry. The significant points here are that
we do not resort to the intuitionally obtuse imaginary-time formalism
and also that our conversion does not necessitate any regularization
scheme. We believe that our calculation would shed light upon the
deeper insight towards the temperature inversion symmetry owing to the
transparency of each procedure and absence of any cutoff added by hand.

Furthermore on occasions one might encounter controversies in
understanding the results acquired at finite temperature and
those at finite extension, or the results in the Euclidean space-time
and those in the Minkowskian space-time. It is often the case that
the topological object existing in the Euclidean world cannot be
interpreted as a physical object in the Minkowskian world from the
thermodynamic point of view (e.g.\ see \cite{bel92}). Also some
subtleties might come from the {\em Wick rotation} from one world to
the other, especially when the theory contains fermionic fields.
Thus it would be informative to see the concrete correspondence
between Planck's law and the Casimir force without resorting to
the $\mathrm{O}(4)$ symmetry of the Euclidean space-time. Our
conclusion will be that the construction of thermodynamics in terms of
the effective temperature (spatial extension) cannot be achieved in
spite of the $\mathrm{O}(4)$ symmetry.

Let us consider the following configuration: there are two square
conducting metal plates with each side sized $L$, one of which is
located at $z=0$ in the $x$-$y$ plane and the other located at $z=l$
parallel to the $x$-$y$ plane, where $l\ll L$. In the present case
where $L$ is regarded as quite large, we can treat the wave numbers
along the $x$ and $y$ directions as continuous. As for the $z$
direction the fixed boundary condition imposed by the metals obliges
the wave number to take the discrete values,
\begin{equation}
 k_z=\frac{\pi n}{l},\qquad n=0,1,2,\dots
\end{equation}
Then noting that the energy quanta of the zero-point oscillation is
given by $\frac{1}{2}\hbar ck$ and that the number of the degrees of
freedom corresponding to the polarization is two except for the zero
mode, we can readily write down the zero-point energy as
\begin{equation}
 U_0=\hbar cL^2\int\frac{\d k_x\d k_y}{(2\pi)^2}\left(\frac{1}{2}
 k_\perp+\sum_{n=1}^{\infty}k_n\right)\equiv\hbar cL^2\int
 \frac{\d k_x\d k_y}{(2\pi)^2}I(k_x,k_y),
\end{equation}
where
\begin{equation}
 k_\perp=\sqrt{k_x^2+k_y^2},\qquad
 k_n=\sqrt{k_\perp^2+\left(\frac{\pi n}{l}\right)^2}.
\end{equation}

Cauchy's integral theorem enables us to rewrite $I(k_x,k_y)$ in the
form of the contour integration in the complex plane,
\begin{equation}
 I(k_x,k_y)=-\oint\frac{\d k}{2\pi}\left(\frac{k^2}{k^2+k_\perp^2}+
 \sum_{n=1}^{\infty}\frac{2k^2}{k^2+k_n^2}\right),
\end{equation}
where the integration contour is a semi-circle with infinitely large
radius whose diameter is on the real axis. As is often performed, we
make use of the formula,
\begin{equation}
 \frac{\coth z}{z}=\frac{1}{z^2}+\sum_{n=1}^{\infty}
 \frac{2}{z^2+\pi^2 n^2},
\end{equation}
to evaluate the summation over the Matsubara frequency, i.e.
\begin{eqnarray}
 I(k_x,k_y) &=& -\oint\frac{\d k}{2\pi}\frac{lk^2}{\sqrt{k^2
  +k_\perp^2}}\coth(l\sqrt{k^2+k_\perp^2})\nonumber\\
 &=&-\oint\frac{\d k}{2\pi}\frac{lk^2}{\sqrt{k^2+k_\perp^2}}-\oint
  \frac{\d k}{2\pi}\frac{2lk^2}{\sqrt{k^2+k_\perp^2}
  (\e^{2l\sqrt{k^2+k_\perp^2}}-1)}.
\label{eq:i-kk}
\end{eqnarray}

Up to now nothing new appears in our procedure to evaluate the
summation \cite{mit72}. Here it will be worth noting again what is
most important in our evaluation is that no term is dropped off in
spite of the presence of apparently divergent (ill-defined) terms (of
course, we can rigorously define them using some proper regularization
only if we do not mind making the expressions a bit more jumbled).
We comment upon that the expression (\ref{eq:i-kk}) mathematically
corresponds to the Abel-Plana formula,
\begin{equation}
 \sum_{n=0}^{\infty}f(n)=\int_0^\infty\d x f(x)+\frac{1}{2}f(0)
 +\mathrm{i}\int_0^\infty\d x\frac{f(\mathrm{i}x)-f(-\mathrm{i}x)}
 {\e^{2\pi x}-1}.
\end{equation}

Let us first consider about the second term since it is more easily
simplified. $\sqrt{k^2+k_\perp^2}$ may take twofold values in the
complex plain. We must specify which one to take the square root
before going on our discussion. Because the integrand of
(\ref{eq:i-kk}) is an even function with respect to
$\sqrt{k^2+k_\perp^2}$, we can choose it in such a way that the
real-part of $\sqrt{k^2+k_\perp^2}$ becomes positive. Then the
contribution from the arched path located infinitely far away vanishes
exponentially. What is left is only the contribution from the path on
the real axis, that can be integrated by parts into (we write $k_z$
here instead of $k$ for later convenience)
\begin{equation}
 -\int_{-\infty}^\infty\frac{\d k_z}{2\pi}\frac{2lk_z^2}
 {\sqrt{k_z^2+k_\perp^2}(\e^{2l\sqrt{k_z^2+k_\perp^2}}-1)} =
 \int_{-\infty}^\infty\frac{\d k_z}{2\pi}\ln(1-\e^{-2l\sqrt{k_z^2
 +k_\perp^2}}).
\label{eq:int_part}
\end{equation}

\begin{figure}[b]
\begin{center}
\begin{picture}(80,70)
\put(0,10){\postscriptbox{8cm}{5cm}{fig1.eps}}
\put(80,12){${\rm Re}k$} \put(41,62){${\rm Im}k$}
\put(46,31){${\rm i}k_\perp$}
\put(32,13){$-\epsilon$} \put(44,13){$+\epsilon$}
\end{picture}
\caption{The modification of the paths in the first and second
quadrants.}
\label{fig:rotation}
\end{center}
\end{figure}
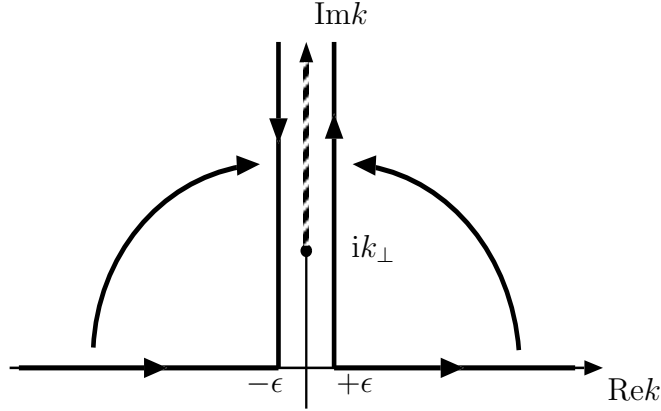

Then we will proceed towards the evaluation of the first term of
(\ref{eq:i-kk}). Since the singularities of the integrand lie on the
imaginary axis, we can modify the path in the first quadrant into the
line apart from the imaginary axis by $+\epsilon$ and the path in the
second quadrant into the line apart from the imaginary axis by
$-\epsilon$ (see Fig.\ \ref{fig:rotation}). The important point is
that the branch cut on the imaginary axis in the region
${\rm Im}k>k_\perp$ corresponds to our way how to specify the sign of
the square root. As a result, the integrations over the region
${\rm Im}k<k_\perp$ cancel out each other and the remaining
integrations result in
\begin{equation}
 -\int_{\mathrm{i}\infty-\epsilon}^{\mathrm{i}k_\perp-\epsilon}
 \frac{\d k}{2\pi}\frac{lk^2}{\sqrt{k^2+k_\perp^2}}-
 \int_{\mathrm{i}k_\perp+\epsilon}^{\mathrm{i}\infty+\epsilon}
 \frac{\d k}{2\pi}\frac{lk^2}{\sqrt{k^2+k_\perp^2}}
 =l\int_{-\infty}^\infty\frac{\d k_z}{2\pi}\sqrt{k_z^2+k_\perp^2},
\end{equation}
where we changed the integration variable from $k$ to
$k_z=-\mathrm{i}\sqrt{k^2+k_\perp^2}$.

Thus without any obscurity the expression (\ref{eq:i-kk}) is
transformed into a considerably concise form, that is
\begin{equation}
 I(k_x,k_y)=l\int_{-\infty}^\infty\frac{\d k_z}{2\pi}k
 +\int_{-\infty}^\infty\frac{\d k_z}{2\pi}\ln(1-\e^{-2lk}).
\end{equation}
Then the zero-point energy is expressed as
\begin{equation}
 U_0=\hbar cL^2l\int\frac{\d^3k}{(2\pi)^3}\left\{k+\frac{1}{l}
 \ln(1-\e^{-2lk})\right\}.
\end{equation}
The finite difference from the continuum counterpart given by
\begin{equation}
 \Delta U_0=U_0-U_0^{\rm cont}=\hbar cL^2\int\frac{\d^3k}{(2\pi)^3}
 \ln(1-\e^{-2lk})
\end{equation}
or the energy density
\begin{equation}
 u=\frac{\Delta U_0}{L^2 l}=\frac{\hbar c}{l}\int\frac{\d^3k}
 {(2\pi)^3}\ln(1-\e^{-2lk})
\end{equation}
generates the Casimir force, that is an attractive force acting on the
plate per unit area,
\begin{equation}
 p=-\frac{\partial(l u)}{\partial l}=-2\hbar c
 \int\frac{\d^3k}{(2\pi)^3}\frac{k}{\e^{2lk}-1}.
\label{eq:casimir}
\end{equation}

On the other hand the thermodynamic functions at finite temperature
$T$ are given by the followings:
\begin{eqnarray}
 p &=&-f=-\frac{2\hbar c}{\beta}\int\frac{\d^3k}{(2\pi)^3}
  \ln(1-\e^{-\beta k}),
\label{eq:pressure}\\
 u &=&f+Ts=f-T\frac{\partial f}{\partial T}
  =-\frac{\partial(\beta p)}{\partial\beta}=2\hbar c\int
  \frac{\d^3k}{(2\pi)^3}\frac{k}{\e^{\beta k}-1},
\label{eq:planck}
\end{eqnarray}
where $p$, $f$, $u$ and $s$ are the pressure, the Helmholtz free
energy density, the internal energy density and the entropy density,
respectively. $\beta$ stands for the inverse temperature
$\beta=\hbar c/k_{\mathrm{B}}T$. The expression of $u$ is nothing but
Planck's law of radiation. Looking at the expressions
(\ref{eq:casimir}) and (\ref{eq:planck}) we can recognize the
{\em duality} relations,
\begin{equation}
 2l\Longleftrightarrow\beta=\frac{\hbar c}{k_{\mathrm{B}}T},\qquad
 p\Longleftrightarrow -u.
\label{eq:corr}
\end{equation}
The reason why the appearance of the additional coefficient in front
of $l$ is that we adopted the fixed boundary condition in the spatial
direction as mentioned at the beginning of this paper. This duality
relation is an embodiment of the symmetry under the exchange of the
temporal axis and the spatial axis, which corresponds to the swap of
the electric field (temporal component) and the magnetic field
(spatial component) in the electrodynamics, that is, the
electro-magnetic duality.

As long as concerned with the $\mathrm{O}(4)$ symmetry, one would
regard $l$ as the (inverse) effective temperature for the system.
As is obvious in the calculation of the partition function in the
functional integral method \cite{wot90}, each mathematical procedure
is absolutely symmetric. Nevertheless physics is different, or,
actually the latter relation in (\ref{eq:corr}) prevents
us from accepting $l$ as the genuine temperature in a thermodynamic
sense. For instance the entropy density in the thermodynamics can be
written in terms of the pressure $p$ and the internal energy density
$u$ as
\begin{equation}
 s=\frac{p+u}{T},
\end{equation}
where the explicit expressions are derived from the equations
(\ref{eq:pressure}) and (\ref{eq:planck}) as
$p=\pi^2\hbar c/45\beta^4$ and $u=\pi^2\hbar c/15\beta^4$. We can
immediately confirm ourselves that the thermodynamic relation,
\begin{equation}
 \frac{\partial s}{\partial u}=\frac{1}{T},
\label{eq:thermo}
\end{equation}
is satisfied, which is the definition of the absolute temperature.
Once we admit the duality relation (\ref{eq:corr}), the dual entropy
density at finite extension is given by
\begin{equation}
 s=-\frac{2k_{\mathrm{B}}l}{\hbar c}(u+p).
\end{equation}
Then the thermodynamic relation becomes
\begin{equation}
 \frac{\partial s}{\partial u}=\frac{6k_{\mathrm{B}}l}{\hbar c}
 \neq \frac{2k_{\mathrm{B}}l}{\hbar c},
\end{equation}
which shows the inconsistency for the thermodynamic relations. In fact
the canonical ensemble in the statistical mechanics is based upon the
relation (\ref{eq:thermo}). The collapse of the relation
(\ref{eq:thermo}) means that we cannot consider any thermodynamic
system in terms of the effective temperature
$\hbar c/2k_{\mathrm{B}}l$.

Thus we have seen the explicit conversion from the Casimir force to
Planck's law of radiation, as is evident in the duality relation
(\ref{eq:corr}). The prominent feature we would like to stress here is
that we could establish the correspondence between the Casimir force
and Planck's law of radiation by resorting neither to the subtle
${\rm O}(4)$ symmetry nor to any artificial regularization. As far as
we know, no one had ever expressed the Casimir force in the form
plainly comparable with Planck's law, like our goal (\ref{eq:planck}).
After the momentum integration for the Casimir force, we reach the
well-known functional form of $l^{-4}$, whose counterpart in the
thermodynamics, of course, is the Stefan-Boltzmann law. What should be
noted here is that the correspondence is between the pressure and the
internal energy density, that have the same dimension. This is the
reason why the system cannot be described in the language of
thermodynamics by using the effective temperature, regardless of the
almost trivial realization of the temperature inversion symmetry in
the framework of the Euclidean functional method owing to the
$\mathrm{O}(4)$ symmetry. We believe that our contribution presented
here will provide an intuitive view in the forefront of physics.

We thank Y.\ Abe for his sincere encouragement to complete our work.

\end{document}